\documentclass[12pt]{article}
\usepackage{graphicx}

\begin{document}
\newcommand  {\ba} {\begin{eqnarray}}
\newcommand  {\be} {\begin{equation}}
\newcommand  {\ea} {\end{eqnarray}}
\newcommand  {\ee} {\end{equation}}

\begin{flushright}
 BIHEP-TH--2002-12
\end{flushright}
\vskip 3.0cm
\begin{center}
{\LARGE\bf CP Asymmetry in Tau Slepton Decay in The Minimal Supersymmetric Standard Model}
 \vskip 1.0cm
{\bf Wei Min Yang\footnote{Email: yangwm@mail.ihep.ac.cn}\hspace{1cm}Dong Sheng Du}\\
 {\em CCAST(World Laboratory), P.O.Box 8730, Beijing 100080, China\\
 Institute of High Energy Physics, P.O.Box 918(4), Beijing 100039, China}
 \vskip 2.0cm
{\Large\bf Abstract}\\
\end{center}

\hspace{20pt} We investigate CP violation asymmetry in the decay of tau slepton into a tau neutrino and a chargino
in the minimal supersymmetric standard model. The new source of CP violation is the complex mixing in the tau
slepton sector. The rate asymmetry between the decays of tau slepton and its CP conjugate process can be of order of
$10^{-3}$ in some region of the parameter space of the mSUGRA scenario, which will possibly be detectable in the
near-future collider experiments.

\newpage

\section*{I. Introduction}
\hspace{20pt}Supersymmetric(SUSY) model is now widely regarded to be the most plausible extension of the Standard
Model(SM)\cite{Nills}\cite{Drees}. It stabilizes the gauge hierarchy and allows for the grand unification of all
known gauge interactions\cite{Witten}. In the Minimal Supersymmetric Standard Model(MSSM) there are additional
complex couplings which may give rise to CP violation compared to the SM\cite{Dimopoulos}. These new sources of CP
violation contribute to CP violation in various processes\cite{Masier}, the neutron and electron electric dipole
moments, rare Kaon decays and B decays, etc. Although the CP-violating phases associated with sfermions of the first
and second generation are severely constrained by bounds on the electric dipole moments of the electron, neutron and
muon, but the CP violation phases in the mass matrices of the third generation sfermions might be large and can
induce sizable CP violation in the MSSM Higgs sector through loop corrections\cite{Nath}\cite{Pilaftsis},
furthermore, these phases also directly affect the couplings of Higgs bosons to third generation
sfermions\cite{Fred}. In the MSSM with the simple universal soft supersymmetry breaking\cite{Chamseddine}, the tau
slepton sector contains two new sources of CP violation in its couplings to Higgs particles. It can be defined to be
the complex phases of the $\mu$ term in the Higgs superpotential and soft-SUSY-breaking \textit{A} terms. These new
sources of CP violation are generic to all SUSY theories and provide non-SM sources of CP violation required for the
baryon asymmetry of the universe\cite{Cline}. On the other hand, the effects of these new CP violation are expected
to be probed in the near-future colliders\cite{Choi}, such as LHC and NLC, which could provide an opportunity of
detecting new CP violating phenomena. Therefore, it should be important and interesting to investigate mechanism and
consequence of this CP violating source in collider phenomenology.

In this paper, we work in the framework of the MSSM with CP violation, and we focus on the CP asymmetry in tau
slepton decays. If a tau slepton is sufficiently heavy, its dominant decay modes are tree-level two-body decays
$\widetilde{\tau}^-\rightarrow\nu_{\tau}\widetilde{\chi}^-$ and
$\widetilde{\tau}^-\rightarrow\tau^-\widetilde{\chi}^0$, where $\widetilde{\chi}^-$ and $\widetilde{\chi}^0$ denote
a chargino and a neutralino, respectively. These final states are also produced at one-loop level by the final state
interactions. If the interactions of the tau slepton violate CP invariance, these decays are expected to have different
rates from their CP conjugate process, which are measured by the asymmetries:\\
\parbox{12.9cm}
{\begin{eqnarray*}
 A_{cp}^{\nu_\tau} &=& \frac{\Gamma(\widetilde{\tau}^-\rightarrow\nu_{\tau}\widetilde{\chi}^-)-
 \Gamma(\widetilde{\tau}^+\rightarrow\bar{\nu}_{\tau}\widetilde{\chi}^+)}
 {\Gamma(\widetilde{\tau}^-\rightarrow\nu_{\tau}\widetilde{\chi}^-)+
 \Gamma(\widetilde{\tau}^+\rightarrow\bar{\nu}_{\tau}\widetilde{\chi}^+)}\:,\\
 A_{cp}^{\tau} &=& \frac{\Gamma(\widetilde{\tau}^-\rightarrow\tau^-\widetilde{\chi}^0)-
 \Gamma(\widetilde{\tau}^+\rightarrow\tau^+\widetilde{\chi}^0)}
 {\Gamma(\widetilde{\tau}^-\rightarrow\tau^-\widetilde{\chi}^0)+
 \Gamma(\widetilde{\tau}^+\rightarrow\tau^+\widetilde{\chi}^0)}\:.
\end{eqnarray*}}
\parbox{0.1cm}{\ba\ea}\\
 Assuming CPT invariance, the decay widths satisfy the relation:
\ba \Gamma(\widetilde{\tau}^-\rightarrow\nu_{\tau}\widetilde{\chi}^-)-
\Gamma(\widetilde{\tau}^+\rightarrow\bar{\nu}_{\tau}\widetilde{\chi}^+)
 =-[\Gamma(\widetilde{\tau}^-\rightarrow\tau^-\widetilde{\chi}^0)-
\Gamma(\widetilde{\tau}^+\rightarrow\tau^+\widetilde{\chi}^0)] \ea
 making the total width of the tau slepton the same as that of the anit-tau slepton. We will calculate and discuss the
 asymmetries.

 The remainder of this paper is organized as follows. Section II is devoted to a brief review of the mass spectra
 and mixing patters of the tau sleptons, charginos and neutralinos. In Sec.III we present a Lagrangian of the relevant
 coupling and analytical expressions of the rate asymmetry. A detailed numerical analysis of the asymmetries for a
 representative SUSY parameter set is given in Sec.IV. Final section is for summary and conclusions.

\section*{II. SUSY particles masses and mixing}
\hspace{20pt}To fix our notation, we simply summarize in this section the masses and mixings of the tau slepton and
chargino, neutralino sectors of the MSSM, which will be needed later when evaluating the decay widths\cite{Gunion}.

\subsection*{A. Tau slepton mass and mixing}
\hspace{20pt}The mass-squared matrix for the tau slepton in the left-right basis is given as
 \ba M^2_{\widetilde{\tau}}=\left(\begin{array}{cc}
     M^2_L & m_\tau(A_\tau + \mu\tan\beta) \\
     m_\tau(A^*_\tau + \mu^*\tan\beta) & M^2_R\end{array}\right)
 \ea
 with\\
\parbox{12.9cm}
{\begin{eqnarray*}
 M^2_L &=& m^2_{\widetilde{\tau}_L}+m^2_\tau+(\frac{1}{2}M^2_Z-M^2_W)\cos2\beta\:,\\
 M^2_R &=& m^2_{\widetilde{\tau}_R}+m^2_\tau+(M^2_Z-M^2_W)\cos2\beta\:,
\end{eqnarray*}}
\parbox{0.1cm}{\ba\ea}\\
 where $m_{\widetilde{\tau}_{L,R}}$ are the left- and right-handed soft SUSY-breaking tau slepton masses, respectively.
 The tau slepton soft breaking trilinear coupling $A_\tau$ and Higgs mass mixing parameter $\mu$ are complex,
\ba A_\tau=|A_\tau|e^{i\,\varphi}\:,\hspace{1cm}\mu_\tau=|\mu_\tau|e^{i\,\eta}\:. \ea
  The complex phase $\varphi$ and $\eta$ are the source of CP violation, which can vary in the range
$0\leq\varphi,\eta\leq2\pi$. In the later context, we will take $\mu$ as real, i.e. only the phase $\varphi$ is left
so as to reduce the number of parameters and simplify the discussion. The tau slepton mass eigenstates can be
realized by a unitary transformation $U$ which diagonalizes the mass-squared matrix $M^2_{\widetilde{\tau}}$\,,
 \ba U\,M^2_{\widetilde{\tau}}\,U^\dagger=Diag\,(m^2_{\widetilde{\tau}_1}\,,\,m^2_{\widetilde{\tau}_2})\,,\ea
 where diagonalization matrix can be parameterized as
 \ba U=\left(\begin{array}{cc}\cos\theta_\tau & \sin\theta_\tau e^{i\,\delta} \\
 -\sin\theta_\tau e^{-i\,\delta} & \cos\theta_\tau\end{array}\right) \ea
 with
 \ba \delta=\arg\,(A_\tau+\mu\tan\beta)\,.\ea
 The tau slepton mixing angles and mass eigenvalues are then given as\\
\parbox{12.9cm}
{\begin{eqnarray*}
\tan2\theta &=& \frac{2m_\tau|A_\tau+\mu\tan\beta|}{M^2_L-M^2_R}\,,\\
m^2_{\widetilde{\tau}_{1,2}} &=&
\frac{1}{2}\left[M^2_L+M^2_R\mp\sqrt{(M^2_L-M^2_R)^2+4m^2_\tau|A_\tau+\mu\tan\beta|^2}\right]
\end{eqnarray*}}
\parbox{0.1cm}{\ba\ea}\\
with convention $0\leq\theta\leq\frac{\pi}{2},\,m^2_{\widetilde{\tau}_1}\leq m^2_{\widetilde{\tau}_2}$. For large
values of $\tan\beta$ and $\mu$, the mixing in the tau slepton sector can be very strong.

\subsection*{B. The chargino and neutralino systems}
\hspace{20pt}The general chargino mass matrix is given by\cite{Djouadi}
 \ba M_C=\left(\begin{array}{cc} M_2 & \sqrt{2}M_W s_\beta \\
 \sqrt{2}M_W c_\beta & \mu\end{array}\right)\,, \ea
 where $M_2$ is the wino mass parameter and we use $s_\beta=\sin\beta,\,c_\beta=\cos\beta$, etc. It can be
diagonalized by two real rotation matrices $C_L$ and $C_R$\,,
 \ba C_L\,M_C\,C_R^{-1}=Diag\,(m_{\widetilde{\chi}^{\pm}_1}\,,\,m_{\widetilde{\chi}^{\pm}_2})\ea
with two rotation angles given by\\
\parbox{12.9cm}
{\begin{eqnarray*}
\tan2\theta_L &=& \frac{2\sqrt{2}M_W(M_2c_\beta+\mu s_\beta)}{M^2_2-\mu^2-2M_W^2c_\beta}\,,\\
\tan2\theta_R &=& \frac{2\sqrt{2}M_W(M_2s_\beta+\mu c_\beta)}{M^2_2-\mu^2+2M_W^2c_\beta}\,.
\end{eqnarray*}}
\parbox{0.1cm}{\ba\ea}\\
The two mass eigenvalues of the charginos, in the limit $|\mu|\gg M_2\,,M_W$\,, are reduced to\\
\parbox{12.9cm}
{\begin{eqnarray*}
m_{\widetilde{\chi}^{\pm}_1} & \simeq & M_2-\frac{M_W^2}{\mu^2}(M_2+\mu s_{2\beta})\,,\\
m_{\widetilde{\chi}^{\pm}_2} & \simeq & |\mu|+\frac{M_W^2}{\mu^2}\epsilon_\mu(M_2 s_{2\beta}+\mu)\,,
\end{eqnarray*}}
\parbox{0.1cm}{\ba\ea}\\
where $\epsilon_\mu$ is for the sign of $\mu$\,. For $|\mu|\rightarrow\infty$\,, the lighter chargino corresponds to
a pure wino state with mass $m_{\widetilde{\chi}^{\pm}_1}\simeq M_2$\,, while the heavier chargino corresponds to a
pure higgsino state with mass $m_{\widetilde{\chi}^{\pm}_2}\simeq |\mu|$\,.

The neutralino mass matrix is
 \ba M_N=\left(\begin{array}{cccc}
     M_1 & 0 & -M_Z s_W c_\beta & M_Z s_W s_\beta \\
     0 & M_2 & M_Z c_W c_\beta & -M_Z c_W s_\beta \\
     -M_Z s_W c_\beta & M_Z c_W c_\beta & 0 & -\mu\\
     M_Z s_W s_\beta & -M_Z c_W s_\beta & -\mu & 0
 \end{array}\right)\,,\ea
where $M_1$ is the Bino mass parameter and $s_W=\sin\theta_W,\,c_W=\cos\theta_W$, etc. are used. It can be
diagonalized by a single real orthogonal matrix $N$\,,
 \ba N\,M_N\,N^{-1}=Diag\,
    (m_{\widetilde{\chi}^0_1}\,,\,m_{\widetilde{\chi}^0_2}\,,\,m_{\widetilde{\chi}^0_3}\,,\,m_{\widetilde{\chi}^0_4})\,.
 \ea
In the limit of large $|\mu|$ values, the mass eigenvalues of the neutralinos
$m_{\widetilde{\chi}^0_i}\,(i=1,2,3,4)$ are simplified to\\
\parbox{12.9cm}
{\begin{eqnarray*}
m_{\widetilde{\chi}^0_1} & \simeq & M_1-\frac{M_Z^2}{\mu^2}(M_1+\mu s_{2\beta})s_W^2\,,\\
m_{\widetilde{\chi}^0_2} & \simeq & M_2-\frac{M_Z^2}{\mu^2}(M_2+\mu s_{2\beta})c_W^2\,,\\
m_{\widetilde{\chi}^0_3} & \simeq &|\mu|+\frac{M_Z^2}{2\mu^2}\epsilon_\mu(1-s_{2\beta})(\mu+M_2 s_W^2+M_1 c_W^2)\,,\\
m_{\widetilde{\chi}^0_4} & \simeq &|\mu|+\frac{M_Z^2}{2\mu^2}\epsilon_\mu(1+s_{2\beta})(\mu-M_2 s_W^2-M_1 c_W^2)\,.
\end{eqnarray*}}
\parbox{0.1cm}{\ba\ea}\\
Again, for $|\mu|\rightarrow\infty$\,, two neutralinos are pure gaugino states with masses
$m_{\widetilde{\chi}^0_1}\simeq M_1\,,m_{\widetilde{\chi}^0_2}\simeq M_2$\,, while the two others are pure higgsino
states with masses $m_{\widetilde{\chi}^0_3}\simeq m_{\widetilde{\chi}^0_4}\simeq |\mu|$\,. The matrix elements of
the diagonalizing matrix, $N_{ij}$ with $i,j=1,\ldots,4$\,, are given by\\
\parbox{12.9cm}
{\begin{eqnarray*}
N_{i1} & = & (1+a_i^2+b_i^2+c_i^2)^{-\frac{1}{2}}\,,\\
N_{i2} & = & N_{i1}\,a_i\,,\\
N_{i3} & = & N_{i1}\,b_i\,,\\
N_{i4} & = & N_{i1}\,c_i
\end{eqnarray*}}
\parbox{0.1cm}{\ba\ea}\\
with\\
\parbox{12.9cm}
{\begin{eqnarray*}
 a_i &=&-\frac{1}{\tan\theta_W}\frac{M_1-\epsilon_i\,m_{\widetilde{\chi}_i^0}}
                                     {M_2-\epsilon_i\,m_{\widetilde{\chi}_i^0}}\,,\\
 b_i &=&\left[\mu\,(M_1-\epsilon_i\,m_{\widetilde{\chi}_i^0})(M_2-\epsilon_i\,m_{\widetilde{\chi}_i^0})\right.\\
     & &\left.-M_Z^2 s_\beta c_\beta [(M_1-M_2) c_W^2+M_2-\epsilon_i\,m_{\widetilde{\chi}_i^0}]\right]\\
     & &\left/\left[M_Z s_W (M_2-\epsilon_i\,m_{\widetilde{\chi}_i^0})(\mu\,c_\beta+\epsilon_i\,m_{\widetilde{\chi}_i^0}\,s_\beta)\right]\right.\,,\\
 c_i &=&\left[-\epsilon_i\,m_{\widetilde{\chi}_i^0}(M_1-\epsilon_i\,m_{\widetilde{\chi}_i^0})(M_2-\epsilon_i\,m_{\widetilde{\chi}_i^0})\right.\\
     & &\left.-M_Z^2 c_\beta^2 [(M_1-M_2) c_W^2+M_2-\epsilon_i\,m_{\widetilde{\chi}_i^0}]\right]\\
     & &\left/\left[M_Z s_W(M_2-\epsilon_i\,m_{\widetilde{\chi}_i^0})(\mu\,c_\beta+\epsilon_i\,m_{\widetilde{\chi}_i^0}\,s_\beta)\right]\right.\,,
\end{eqnarray*}}
\parbox{0.1cm}{\ba\ea}\\
where $\epsilon_1=\epsilon_2=1,-\epsilon_3=\epsilon_4=\epsilon_\mu$\,.

\subsection*{C. Mass spectra in the mSUGRA scenario}
\hspace{20pt}To reduce  the number of the parameters, we will adopt the mSUGRA scenario with universality hypothesis
to discuss the SUSY particle spectra, where the scalar fermion masses and the gaugino masses are respectively
unified as $m_0$ and $m_{1/2}$ at the GUT scale $M_{GUT}$\,. The relation between the SUSY particle masses at the
scale $M_{GUT}$ and
at the weak scale ${\cal O}(M_Z)$ are obtained by running renormalization group equations(RGE) as\cite{Boer}\\
\parbox{12.9cm}
{\begin{eqnarray*}
M_1 & \simeq & 0.4\, m_{1/2}\,,\hspace{0.5cm}M_2\,\simeq \,0.8\, m_{1/2}\,,\\
m^2_{\widetilde{\tau}_R} & \simeq & m_0^2+0.15\, m_{1/2}^2-0.23 M_Z^2 \cos2\beta \,,\\
m^2_{\widetilde{\tau}_L} & \simeq & m_0^2+0.52\, m_{1/2}^2-0.27 M_Z^2 \cos2\beta \,,\\
m^2_{\widetilde{\nu}_\tau} & \simeq & m_0^2+0.52\, m_{1/2}^2+0.5 M_Z^2 \cos2\beta \,,
\end{eqnarray*}}
\parbox{0.1cm}{\ba\ea}\\
where $m_{\widetilde{\nu}_\tau}$ is the left-handed soft-SUSY-breaking tau sneutrino mass. All of the free
parameters now include $|A|,\,\varphi,\,\mu,\,\tan\beta,\,m_0,\,m_{1/2}$\,. We will take them as input and use the
above-mentioned equations, all the mass spectra of the involved SUSY particles can then be worked out.

\section*{III. Relevant couplings and decay rate asymmetry}
\begin{figure}[t]
\centering
\includegraphics{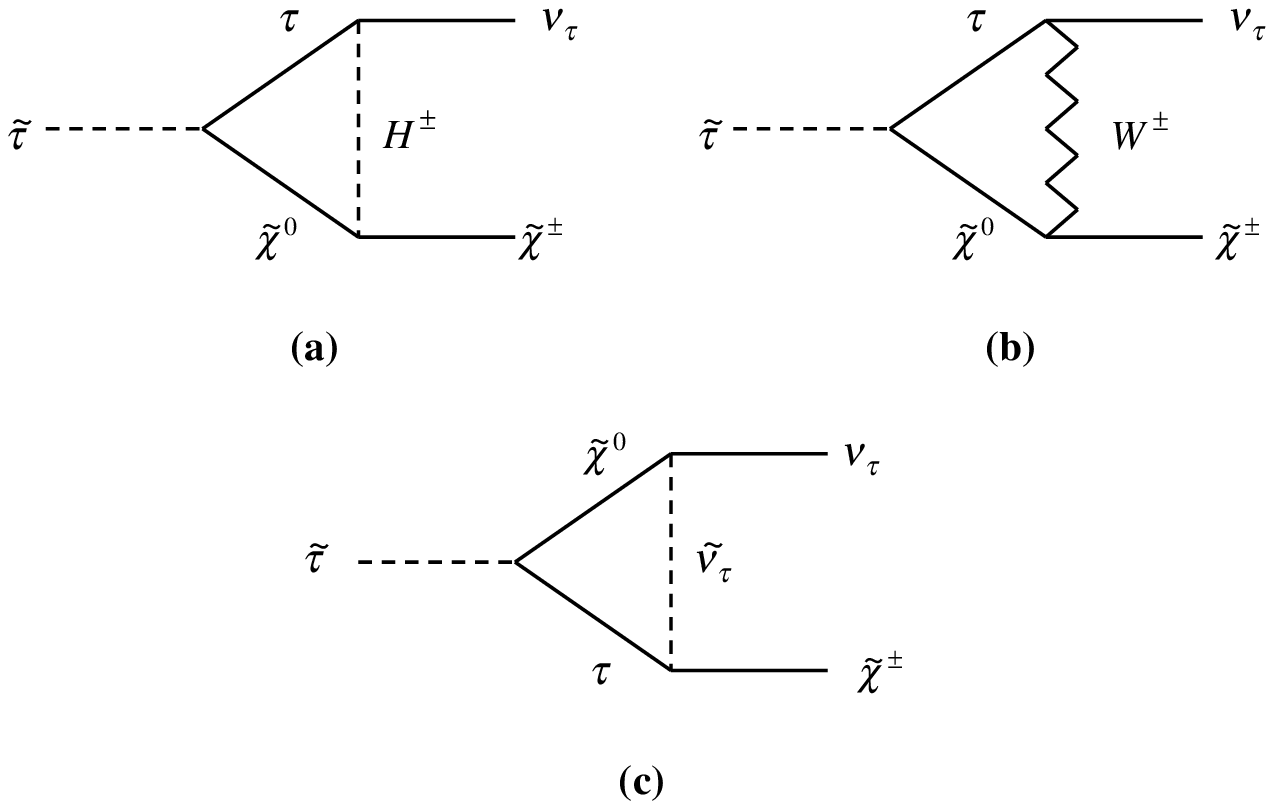} \caption{\it The one-loop diagrams for the decay of a tau slepton into a
tau neutrino and a chargino.\label{loop}}
\end{figure}

\hspace{20pt}A nonvanishing value for the asymmetry $A_{cp}^{\nu_\tau}$ in Eq.(1) is generated, if the decay
$\widetilde{\tau}^-\rightarrow\nu_\tau\widetilde{\chi}^-$, in addition to the decay
$\widetilde{\tau}^-\rightarrow\tau^-\widetilde{\chi}^0$, is allowed kinematically. The produced tau lepton and
neutralino can become a tau neutrino and a chargino by exchanging charged Higgs bosons $H^{\pm}$, $W^{\pm}$ bosons,
and tau sneutrino $\widetilde{\nu}_\tau$, as shown in Fig.~\ref{loop}. The interferences of these one-loop diagrams
with the tree diagram make the rate of the decay $\widetilde{\tau}^-\rightarrow\nu_\tau\widetilde{\chi}^-$ different
from that of the decay $\widetilde{\tau}^+\rightarrow\bar{\nu}_\tau\widetilde{\chi}^+$. The relevant interaction
Lagrangian for $\widetilde{\tau}$ and $\widetilde{\nu}_\tau$ as well as $H^{\pm}$ and $W^{\pm}$ is given by\cite{Gunion}\\

\parbox{12.9cm}
{\begin{eqnarray*} {\cal L}
  &=&\frac{g}{\sqrt{2}}\widetilde{\nu}_\tau\overline{\tau}(A_{Ll}\frac{1-\gamma^5}{2}+A_{Rl}\frac{1+\gamma^5}{2})\widetilde{\chi}_l^-
    +\frac{g}{\sqrt{2}}\widetilde{\tau}_k\overline{\nu_\tau}(B_{kl}\frac{1+\gamma^5}{2})\widetilde{\chi}_l^+ \\
  &&+\frac{g}{\sqrt{2}}\widetilde{\tau}_k\overline{\tau}(F_{Ljk}\frac{1-\gamma^5}{2}+F_{Rjk}\frac{1+\gamma^5}{2})\widetilde{\chi}_j^0
    +\frac{g}{\sqrt{2}}\widetilde{\nu}_\tau\overline{\nu_\tau}(G_j\frac{1+\gamma^5}{2})\widetilde{\chi}_j^0 \\
  &&-\frac{g}{\sqrt{2}}H^-\overline{\widetilde{\chi}_j^0}\gamma^\mu(I_{Ljl}\frac{1-\gamma^5}{2}+I_{Rjl}\frac{1+\gamma^5}{2})\widetilde{\chi}_l^+
    -\frac{g}{\sqrt{2}}H^-\overline{\tau}(K\frac{1-\gamma^5}{2})\nu_\tau \\
  &&-\frac{g}{\sqrt{2}}W_\mu^-\overline{\widetilde{\chi}_j^0}\gamma^\mu(H_{Ljl}\frac{1-\gamma^5}{2}+H_{Rjl}\frac{1+\gamma^5}{2})\widetilde{\chi}_l^+
    -\frac{g}{\sqrt{2}}W_\mu^-\overline{\tau}\gamma^\mu\frac{1-\gamma^5}{2}\nu_\tau \\
  &&+h.c.
\end{eqnarray*}}
\parbox{0.1cm}{\ba\ea}\\
with\\
\parbox{12.9cm}
{\begin{eqnarray*}
  A_{Ll}&=&\frac{m_\tau}{M_W c_\beta}C_{Ll2}\,,\hspace{0.5cm} A_{Rl}=-\sqrt{2}C_{Rl1}\,,\\
  B_{kl}&=&\frac{m_\tau}{M_W c_\beta}C_{Ll2}U_{k2}^*-\sqrt{2}C_{Ll1}U_{k1}^* \,,\\
  F_{Ljk}&=&-\frac{m_\tau}{M_W c_\beta}N_{j3}U_{k1}^*-2\tan\theta_W N_{j1}U_{k2}^* \,,\\
  F_{Rjk}&=&\frac{m_\tau}{M_W c_\beta}N_{j3}U_{k2}^*-(N_{j2}+\tan\theta_W N_{j1})U_{k1}^* \,,\\
  G_j&=&N_{j2}-\tan\theta_W N_{j1}\,,\\
  H_{Ljl}&=&-\sqrt{2}N_{j2}C_{Rl1}+N_{j4}C_{Rl2}\,,\\
  H_{Rjl}&=&-\sqrt{2}N_{j2}C_{Ll1}-N_{j3}C_{Ll2}\,,\\
  I_{Ljl}&=&\cos\beta[\sqrt{2}N_{j4}C_{Rl1}+(N_{j2}+\tan\theta_W N_{j1})C_{Rl2}]\,,\\
  I_{Rjl}&=&\sin\beta[\sqrt{2}N_{j3}C_{Ll1}-(N_{j2}+\tan\theta_W N_{j1})C_{Ll2}]\,,\\
  K&=&-\tan\beta\frac{m_\tau}{M_W}\,,
\end{eqnarray*}}
\parbox{0.1cm}{\ba\ea}\\
where $k,l=(1,2)$ and $j=(1,\ldots,4)$. We now consider the decay of the heavier tau slepton into the tau neutrino
and the lighter chargino $\widetilde{\tau}_2^-\rightarrow\nu_\tau\widetilde{\chi}_1^-$ and its CP conjugate process.
The decay rate asymmetry for the processes are obtained as
 \ba A_{cp}^{\nu_\tau}=\frac{\alpha_2}{2}\frac{T^a+T^b+T^c}{(m^2_{\widetilde{\tau}_2}-m^2_{\widetilde{\chi}_1^-})^2|B_{21}|^2}\,,
 \ea
where $\alpha_2=g^2/(4\pi)$. The contributions of the diagrams (a),(b), and (c) in Fig.1. are represented by $T^a$,
$T^b$, and $T^c$, respectively. these terms are written as\\
\parbox{12.9cm}
{\begin{eqnarray*}
  T^a &=& \sum_j\sum_{n=1}^4 \mbox{Im}(X_j^n)J_n^a(m_{\widetilde{\tau}_2},m_{\widetilde{\chi}_j^0},m_{\widetilde{\chi}_1^-})\,,\\
  T^b &=& \sum_j\sum_{n=1}^4 \mbox{Im}(Y_j^n)J_n^b(m_{\widetilde{\tau}_2},m_{\widetilde{\chi}_j^0},m_{\widetilde{\chi}_1^-})\,,\\
  T^c &=& \sum_j\sum_{n=1}^4 \mbox{Im}(Z_j^n)J_n^c(m_{\widetilde{\tau}_2},m_{\widetilde{\chi}_j^0},m_{\widetilde{\chi}_1^-})
\end{eqnarray*}}
\parbox{0.1cm}{\ba\ea}\\
with \\
\parbox{12.9cm}
{\begin{eqnarray*}
 & X_j^1=KB_{21}F_{Rj2}^*I_{Lj1},& X_j^2=KB_{21}F_{Lj2}^*I_{Lj1},\\
 & X_j^3=KB_{21}F_{Lj2}^*I_{Rj1},& X_j^4=KB_{21}F_{Rj2}^*I_{Rj1};\\
 & Y_j^1=B_{21}F_{Rj2}^*H_{Lj1}, & Y_j^2=B_{21}F_{Rj2}^*H_{Rj1},\\
 & Y_j^3=B_{21}F_{Lj2}^*H_{Lj1}, & Y_j^4=B_{21}F_{Lj2}^*H_{Rj1};\\
 & Z_j^1=A_{R1}B_{21}F_{Rj2}^*G_j, & Z_j^2=A_{R1}B_{21}F_{Lj2}^*G_j,\\
 & Z_j^3=A_{L1}B_{21}F_{Lj2}^*G_j, & Z_j^4=A_{L1}B_{21}F_{Rj2}^*G_j
\end{eqnarray*}}
\parbox{0.1cm}{\ba\ea}\\
and\\
\parbox{12.9cm}
{\begin{eqnarray*}
 J_1^a &=& \frac{1}{2} m_\tau m_{\widetilde{\chi}_1^-}
           \left[T +(m^2_\tau+m^2_{\widetilde{\chi}_1^-}-m^2_{\widetilde{\tau}_2}-m^2_{H^{\pm}})
           \ln \left|\frac{S+T}{S-T}\right|\right],\\
 J_2^a &=& \frac{1}{2} m_{\widetilde{\chi}_j^0} m_{\widetilde{\chi}_1^-}
           \left[T +(m^2_\tau-m^2_{H^{\pm}})\ln \left|\frac{S+T}{S-T}\right|\right],\\
 J_3^a &=& \frac{1}{2}\left[m^2_{\widetilde{\tau}_2} T
           +(m^2_\tau m^2_{\widetilde{\chi}_1^-}-m^2_{\widetilde{\tau}_2} m^2_{H^{\pm}})
           \ln \left|\frac{S+T}{S-T}\right|\right],\\
 J_4^a &=& \frac{1}{2} m_\tau m_{\widetilde{\chi}_j^0}(m^2_{\widetilde{\chi}_1^-}-m^2_{\widetilde{\tau}_2})
           \ln \left|\frac{S+T}{S-T}\right|,\\
 J_1^b &=& \frac{1}{2} m_{\widetilde{\chi}_j^0} m_{\widetilde{\chi}_1^-}
           \left[\frac{2M^2_W+m^2_\tau}{M^2_W}T +(m^2_\tau-2M^2_W+\frac{m^4_\tau}{M^2_W})
           \ln \left|\frac{S^\prime+T}{S^\prime-T}\right|\right],\\
 J_2^b &=& -\frac{m^2_\tau m^2_{\widetilde{\chi}_1^-}}{2M^2_W}T +\left[\frac{1}{2}m^2_\tau m^2_{\widetilde{\chi}_1^-}
         +(m^2_{\widetilde{\chi}_1^-}-m^2_{\widetilde{\tau}_2})(m^2_{\widetilde{\tau}_2}-m^2_\tau-m^2_{\widetilde{\chi}_j^0})\right.\\
      &&+\left.\frac{m^2_\tau m^2_{\widetilde{\chi}_1^-}}{2M^2_W}(m^2_{\widetilde{\chi}_j^0}-m^2_{\widetilde{\chi}_1^-}-m^2_\tau)
         +\frac{m^2_\tau m^2_{\widetilde{\tau}_2}}{2M^2_W}(m^2_{\widetilde{\chi}_1^-}-m^2_{\widetilde{\chi}_j^0})\right]
         \ln \left|\frac{S^\prime+T}{S^\prime-T}\right|,\\
 J_3^b &=& \frac{1}{2} m_\tau m_{\widetilde{\chi}_1^-} \left\{\frac{2M^2_W-m^2_{\widetilde{\tau}_2}+m^2_{\widetilde{\chi}_j^0}}{M^2_W}T
         +\left[ 2(m^2_\tau+m^2_{\widetilde{\chi}_1^-}-M^2_W) \right.\right. \\
      &&-\left.\left. m^2_{\widetilde{\tau}_2}-m^2_{\widetilde{\chi}_j^0}-\frac{m^2_\tau(m^2_{\widetilde{\chi}_1^-}-m^2_{\widetilde{\chi}_j^0})}{M^2_W}\right]
         \ln \left|\frac{S^\prime+T}{S^\prime-T}\right|\right\},\\
 J_4^b &=&\frac{m_\tau m_{\widetilde{\chi}_j^0}(m^2_{\widetilde{\tau}_2}-m^2_{\widetilde{\chi}_1^-})}{2M^2_W}
          \left(T +3M^2_W \ln \left|\frac{S^\prime+T}{S^\prime-T}\right|\right),
\end{eqnarray*}}
\parbox{0.1cm}{\ba\ea}\\
where\\
\parbox{12.9cm}
{\begin{eqnarray*}
 T &=& \frac{(m^2_{\widetilde{\tau}_2}-m^2_{\widetilde{\chi}_1^-})}{m^2_{\widetilde{\tau}_2}}
       \sqrt{m^4_{\widetilde{\tau}_2}+m^4_\tau+m^4_{\widetilde{\chi}_j^0}
       -2m^2_{\widetilde{\tau}_2}m^2_\tau-2m^2_{\widetilde{\tau}_2}m^2_{\widetilde{\chi}_j^0}-2m^2_\tau m^2_{\widetilde{\chi}_j^0}}\,,\\
 S &=& \frac{1}{m^2_{\widetilde{\tau}_2}}(m^2_{\widetilde{\tau}_2}+m^2_{\widetilde{\chi}_1^-})(m^2_{\widetilde{\tau}_2}-m^2_\tau+m^2_{\widetilde{\chi}_j^0})
       -2(m^2_{\widetilde{\chi}_j^0}+m^2_{\widetilde{\chi}_1^-}-m^2_{H^{\pm}})\,,
\end{eqnarray*}}
\parbox{0.1cm}{\ba\ea}\\
$S^\prime$ is derived from $S$ by changing $m_{H^{\pm}}$ to $M_W$. In addition, $J_n^c$ are also obtained from
$J_n^a$ by $m_{\widetilde{\nu}_\tau}$ replacing $m_{H^{\pm}}$. The sum for the intermediate neutralinos in the
formula (23) should be done for those which satisfy the kinematical condition
$m_{\widetilde{\tau}_2}>m_\tau+m_{\widetilde{\chi}_j^0}$ .

\section*{IV. Numerical results}
\hspace{20pt}In this section, we will illustrate our numerical results of the CP asymmetry in the tau slepton decay
based on the mSUGRA scenario for the relevant SUSY parameters. Since we have assumed a universal mass $m_0$ for the
scalar fermions and a mass $m_{1/2}$ for the gauginos at the GUT scale, therefore, the parameters appearing in our
analyses are $|A|,\,\varphi,\,\mu,\,\tan\beta,\,m_0,\,m_{1/2},\,m_{H^{\pm}}$. For simplicity, although these
parameters are not all independent of each other, we assume them are independent and assume only rough constraints
coming from theoretical and experimental considerations. The simple expressions (19) will be used for the soft
SUSY-breaking Bino and Wino mass, as well as left- and right-handed slepton masses when performing the RGE evolution
to weak scale at one-loop order if the Yukawa couplings in the RGE's are neglected. We will choose two
representative values for $\tan\beta$: a low value ($\tan\beta=2.5$) and a large value ($\tan\beta=40$), as well as
two values for phase $\varphi$: $\pi/2$ and $\pi/4$, respectively. The other parameters are typically taken, for
example, as the followings
 \ba \eta=0,\,|\mu|=2 \mbox{TeV},\,|A_\tau|=1.5 \mbox{TeV},\,m_{H^{\pm}}=1 \mbox{TeV},\,m_0=m_{1/2}=400 \mbox{GeV}\,.\ea
As a result, the masses of the relevant SUSY particles are immediately leaded to
 \ba m_{\widetilde{\chi}_1^\pm}\simeq m_{\widetilde{\chi}_2^0}\simeq M_2\simeq 320\mbox{GeV},\,
 m_{\widetilde{\chi}_1^0}\simeq M_1\simeq 160\mbox{GeV},\,m_{\widetilde{\nu}_\tau}\simeq 490\mbox{GeV}\,, \ea
which do not depend on the value of $\varphi$ and there is only a very small variation with the value of
$\tan\beta$. The masses of the tau sleptons , however, are more sensitive to $\tan\beta$, and there is a very small
change with the value of $\varphi$. For $\varphi=\pi/4$, the numerical results is given by\\
\parbox{12.9cm}
{\begin{eqnarray*}
 m_{\widetilde{\tau}_1}\simeq 430\mbox{GeV}\,,& m_{\widetilde{\tau}_2}\simeq 498\mbox{GeV}\,,& (\tan\beta=2.5)\\
 m_{\widetilde{\tau}_1}\simeq 266\mbox{GeV}\,,& m_{\widetilde{\tau}_2}\simeq 604\mbox{GeV}\,.& (\tan\beta=40)
\end{eqnarray*}}
\parbox{0.1cm}{\ba\ea}\\
Since for tau slepton, large enough off-diagonal elements of the mass matrices are obtained only for large $\mu$ and
$\tan\beta$ values and trilinear couplings $A_\tau$ play only a marginal role, we will fix the latter in the entire
analysis. The soft SUSY-breaking masses for the Higgs bosons are however disconnected from the sfermions, moreover,
our results are not sensitive to the charged Higgs bosons mass $m_{H^{\pm}}$, so it will also be fixed.

\begin{figure}[t]
\centering
\includegraphics{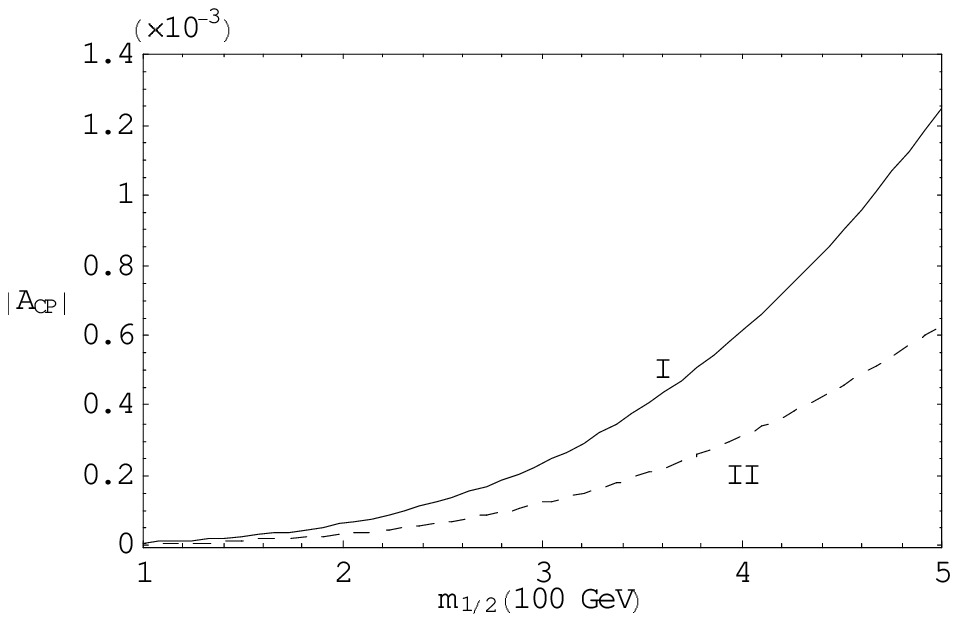}
\caption{\it The decay rate asymmetry as a function of the gaugino mass $m_{1/2}$ for $\tan\beta=2.5$, and the other
parameter values in Eq.(27). The curve {\rm I} $\Rightarrow$ $\varphi=\pi/2$, {\rm II} $\Rightarrow$
$\varphi=\pi/4$.\label{Acp1}}
 \vskip 3.5cm
\includegraphics{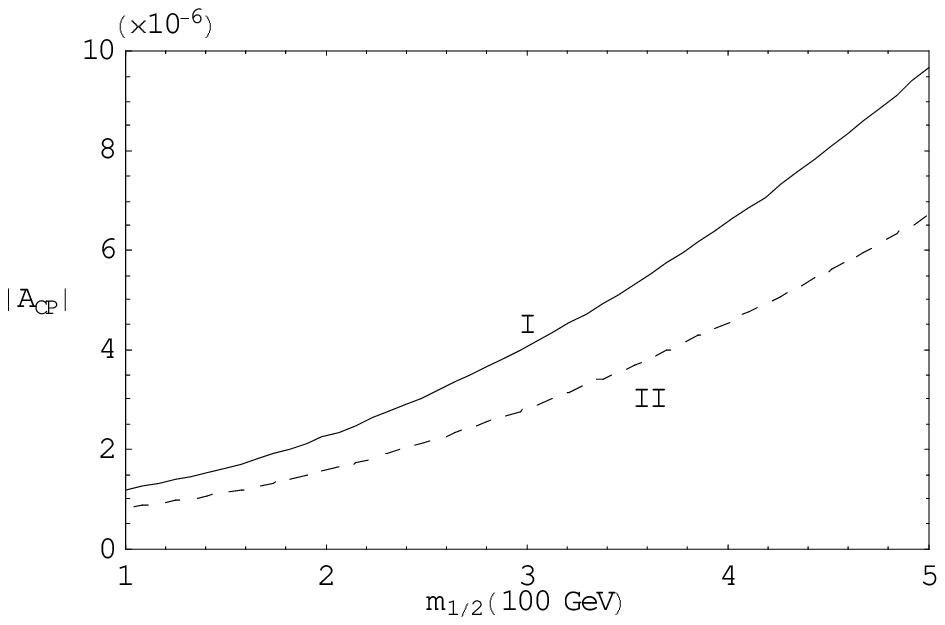}
\caption{\it The decay rate asymmetry as a function of the gaugino mass $m_{1/2}$ for $\tan\beta=40$, and the other
parameter values in Eq.(27). The curve {\rm I} $\Rightarrow$ $\varphi=\pi/2$, {\rm II} $\Rightarrow$
$\varphi=\pi/4$.\label{Acp2}}
\end{figure}

In the Fig.~\ref{Acp1} and Fig.~\ref{Acp2}, the absolute values of $A_{cp}^{\nu_\tau}$ are shown as a function of
the unified gaugino mass $m_{1/2}$ for the low value ($\tan\beta=2.5$) and the large value ($\tan\beta=40$),
respectively. The other involved parameters are fixed to the same as Eq.(27). Two curves of each figure correspond
respectively to two value of the phase $\varphi=\pi/4$ and $\varphi=\pi/2$. The plots show that the rate asymmetries
$|A_{cp}^{\nu_\tau}|$ are very sensitive to the value of $\tan\beta$. They have approximately a magnitude of order
of $10^{-4}$ for $\tan\beta=2.5$ and of order of $10^{-6}$ for $\tan\beta=40$, respectively. The asymmetries are
enhanced with increasing value of $m_{1/2}$. In the case of $\tan\beta=2.5$ and $\varphi=\pi/2$, for large values of
$m_{1/2}$($m_{1/2}\approx 500\mbox{GeV}$), the asymmetry $A_{cp}^{\nu_\tau}$ can significantly reach the order of
$10^{-3}$. In addition, the parameters $\varphi,\,|A_\tau|,\,m_0$ do dot change the whole trends of the plots,
nevertheless, they can slightly shift the values of $A_{cp}^{\nu_\tau}$ in the same order of magnitude. The mass of
the charged Higgs boson does not however affect the asymmetries obviously.

\begin{figure}[t]
\centering
\includegraphics{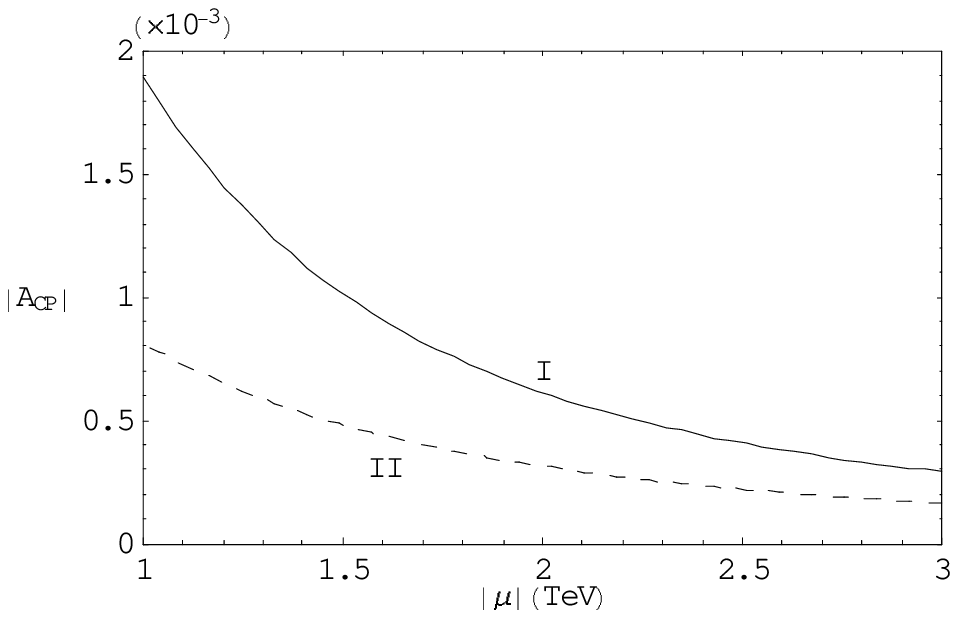}
\caption{\it The decay rate asymmetry as a function of the parameter $|\mu|$ for $\tan\beta=2.5$, and the other
parameter values in Eq.(27). The curve {\rm I} $\Rightarrow$ $\varphi=\pi/2$, {\rm II} $\Rightarrow$
$\varphi=\pi/4$.\label{Acp3}}
 \vskip 3.5cm
\includegraphics{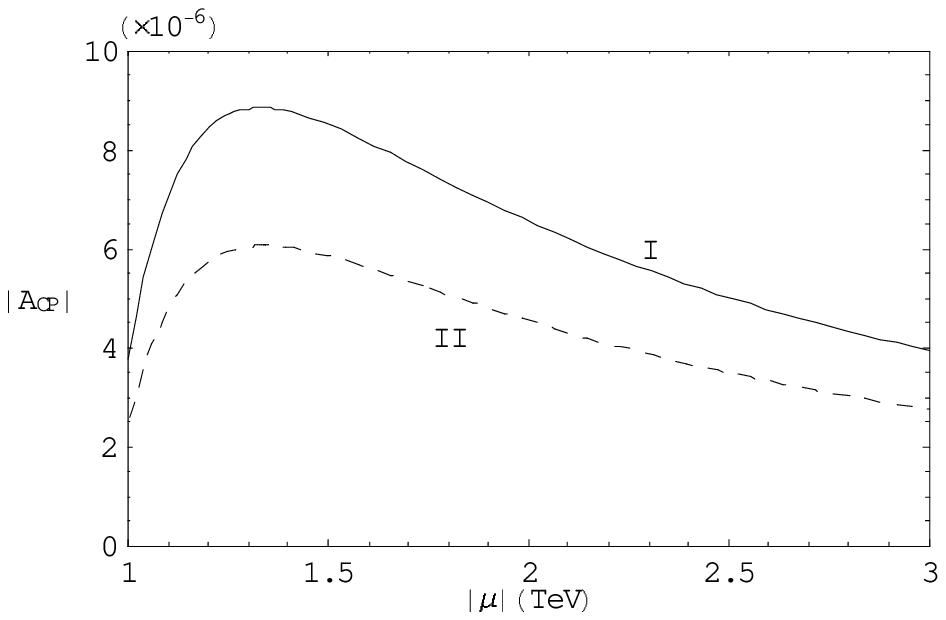}
\caption{\it The decay rate asymmetry as a function of the parameter $|\mu|$ for $\tan\beta=40$, and the other
parameter values in Eq.(27). The curve {\rm I} $\Rightarrow$ $\varphi=\pi/2$, {\rm II} $\Rightarrow$
$\varphi=\pi/4$.\label{Acp4}}
\end{figure}

\begin{figure}[t]
\centering
\includegraphics{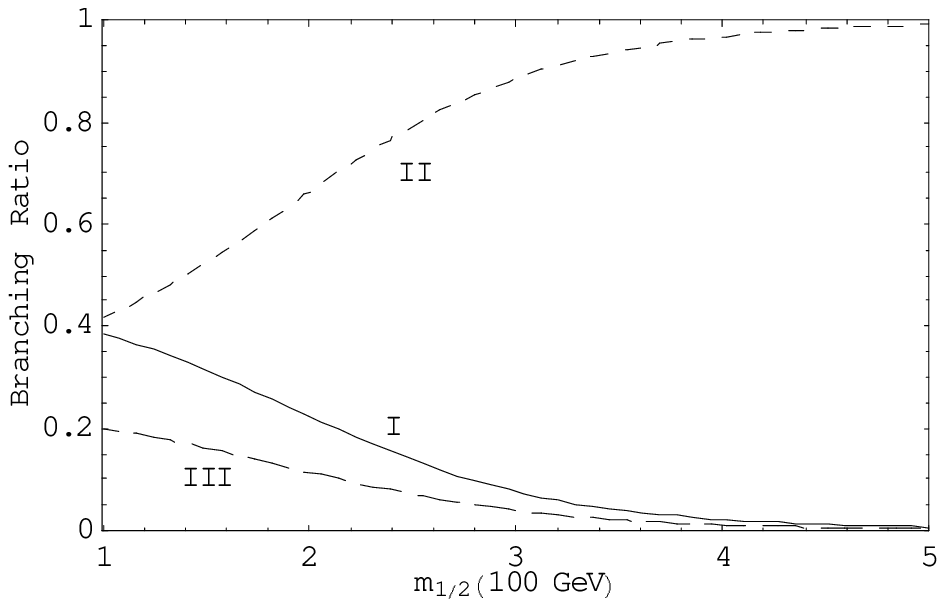}
\caption{\it The branching ratios as a function of the gaugino mass $m_{1/2}$ for $\tan\beta=2.5$, $\varphi=\pi/4$,
and the other parameter values in Eq.(27). The curve {\rm I} $\Rightarrow$
$(\widetilde{\tau_2}\rightarrow\nu_\tau\widetilde{\chi}_1^-)$, {\rm II} $\Rightarrow$
$(\widetilde{\tau_2}\rightarrow\tau\widetilde{\chi}_1^0)$, {\rm III} $\Rightarrow$
$(\widetilde{\tau_2}\rightarrow\tau\widetilde{\chi}_2^0)$.\label{Br1}}
 \vskip 3.5cm
\includegraphics{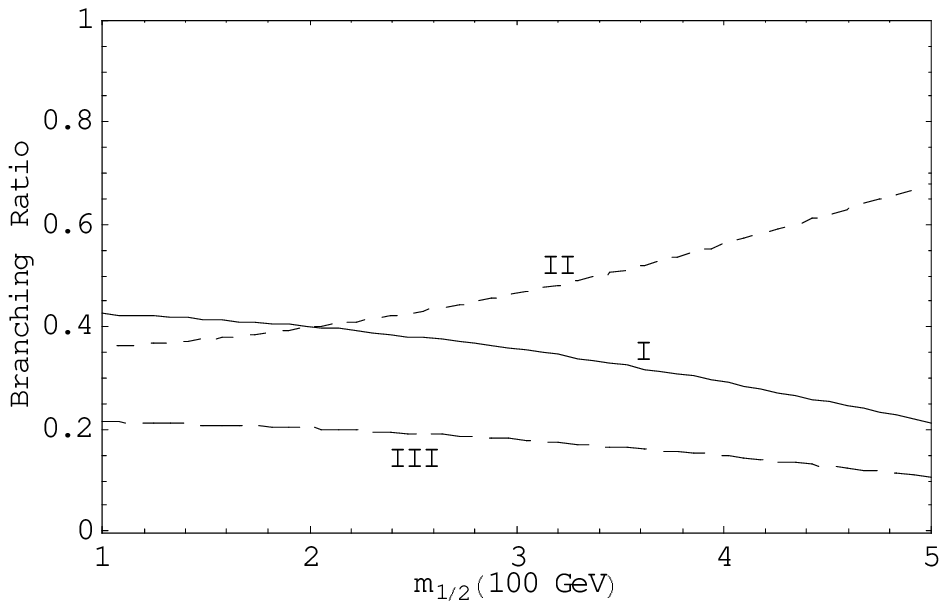}
\caption{\it The branching ratios as a function of the gaugino mass $m_{1/2}$ for $\tan\beta=40$, $\varphi=\pi/4$,
and the other parameter values in Eq.(27). The curve {\rm I} $\Rightarrow$
$(\widetilde{\tau_2}\rightarrow\nu_\tau\widetilde{\chi}_1^-)$, {\rm II} $\Rightarrow$
$(\widetilde{\tau_2}\rightarrow\tau\widetilde{\chi}_1^0)$, {\rm III} $\Rightarrow$
$(\widetilde{\tau_2}\rightarrow\tau\widetilde{\chi}_2^0)$.\label{Br2}}
\end{figure}

In the Fig.~\ref{Acp3} and Fig.~\ref{Acp4}, the rate asymmetries $|A_{cp}^{\nu_\tau}|$ are shown as a function of
the Higgs mass parameter $|\mu|$ for $\tan\beta=2.5$ and $\tan\beta=40$, respectively. The unified gaugino mass and
the other parameters are still given by Eq.(27). Plots show that the magnitude of order of the asymmetries are
similar to that of the Fig.~\ref{Acp1} and Fig.~\ref{Acp2}, respectively. For a larger value of $|\mu|$, the
asymmetries are smaller. In the case of $\tan\beta=2.5$ and $\varphi=\pi/2$, the values of $|A_{cp}^{\nu_\tau}|$
also become of order of $10^{-3}$ for small values of $|\mu|$($|\mu|\approx 1\mbox{TeV}$).
In most region of parameter space in Fig.~\ref{Acp1}--Fig.~\ref{Acp4}, $\widetilde{\tau_2}$ dominantly decays into
$\nu_\tau\widetilde{\chi}_1^-$, $\tau\widetilde{\chi}_1^0$ and $\tau\widetilde{\chi}_2^0$\,. In Fig.~\ref{Br1} and
Fig.~\ref{Br2}, the branching ratios of these decays are shown as a function of the gaugino mass $m_{1/2}$ for
$\tan\beta=2.5$ and $\tan\beta=40$, respectively. Where, the phase $\varphi$ is fixed to $\varphi=\pi/4$ and values
of the other parameters are the same as Eq.(27). The graphs show that the trends of plots obviously change with the
value of $\tan\beta$. For a smaller value of $m_{1/2}$,
$\mbox{Br}(\widetilde{\tau_2}\rightarrow\nu_\tau\widetilde{\chi}_1^-)$ is larger. The interactions which induce the
rate asymmetry between the decays $\widetilde{\tau}^-\rightarrow\nu_\tau\widetilde{\chi}^-$ and
$\widetilde{\tau}^+\rightarrow\bar{\nu}_\tau\widetilde{\chi}^+$ also yield a rate asymmetry between the decays
$\widetilde{\tau}^-\rightarrow\tau^-\widetilde{\chi}^0$ and $\widetilde{\tau}^+\rightarrow\tau^+\widetilde{\chi}^0$,
satisfying the relation in Eq.(3). As seen In Fig.~\ref{Br1} and Fig.~\ref{Br2}, the width of
$\widetilde{\tau}_2\rightarrow\nu_\tau\widetilde{\chi}_1^-$ is generally several times smaller than that of
$\widetilde{\tau}_2\rightarrow\tau\widetilde{\chi}_1^0$, and accordingly the former decay rate asymmetry becomes
larger than the latter by the same order of magnitude. For the detection of an asymmetry
$A_{cp}^{\nu_\tau}\sim10^{-3}$, a necessary number of pairs of $\widetilde{\tau}^+\widetilde{\tau}^-$ should be in
the order of $10^6$. This luminosity are expected to be produced at a future $\mu^+\mu^-$ linear collider with a
c.m. energy of 500GeV, where it will be possible to examine CP violation through the decay
$\widetilde{\tau}\rightarrow\nu_\tau\widetilde{\chi}^\pm$.

\section*{V. Conclusions}
\hspace{20pt}In summary, we have studied in detail the signal for CP violation in the tau slepton sector in the
MSSM. The relevant sources of CP violation come from the soft SUSY-breaking terms associated with third generation
slepton, as well as the Higgs mass parameter $\mu$. We presented a general formalism of the effect of the
CP-violating mixing in the tau slepton sector on their decays. A detailed analysis about that was focused on the
rate asymmetry of the decay of the heavier tau slepton into the lighter chargino and tau neutrino final states. In
the mSUGRA scenarios where the scalar fermion and gaugino masses are unified at the GUT scale, we illustrated this
asymmetry and branching ratios in the parameter space which are constrained by experiments. It was shown that a rate
asymmetry between the decays $\widetilde{\tau}_2^-\rightarrow\nu_\tau\widetilde{\chi}_1^-$ and
$\widetilde{\tau}_2^+\rightarrow\bar{\nu}_\tau\widetilde{\chi}_1^+$ can be induced at a magnitude of order of
$10^{-3}$ in a region of the parameter space where CP violation becomes maximal at lagrangian level. Even though
this CP-violating tau slepton mixing only proceeds through loop diagrams, it can give rise to order of $10^{-3}$
CP-violating asymmetry even in the absence of other CP phases. As a result, the intrinsic property of CP violation
in the MSSM can be expected to be detectable in the near-future collider experiments.

\section*{Acknowledgments}
\hspace{20pt}W.M.Yang thanks M.Ablikim for helpful discussions. This work is in part supported by National Natural
Science Foundation of China.

\end{document}